# Magnetophononics and the Chiral Phonon Misnomer

R. Merlin

*The Harrison M. Randall Laboratory of Physics,*
*University of Michigan, Ann Arbor, Michigan 48109-1040, USA*

The direct, ultrafast excitation of polar phonons with electromagnetic radiation is a potent strategy for controlling the properties of a wide range of materials, particularly in the context of influencing their magnetic behavior. Here, we show that, contrary to common perception, the origin of phonon-induced magnetic activity does not stem from the motion of ions themselves; instead, it arises from the effect their motion exerts on the electron subsystem. Through the mechanism of electron-phonon coupling, a coherent state of circularly polarized phonons generates substantial non-Maxwellian fields that disrupt time reversal symmetry, effectively emulating the behavior of authentic magnetic fields. Notably, the effective field can reach magnitudes as high as 100 T, surpassing by several orders of magnitude the Maxwellian field resulting from the circular motion of the ions. Because the light-induced non-reciprocal fields depend on the square of the phonon displacements, the chirality the photons transferred to the ions plays no role in magnetophononics.

The term helical- or chiral-phonon made its appearance in the literature in the mid-2010s, with the exploration of topological materials [1] and valley-phonon modes in monolayer transition-metal dichalcogenides [2,3]. In its strictest sense, an object is called chiral if it cannot be mapped to its mirror image by rotations and translations alone [4]. However, in the context of phonons, this term has transcended its usual definition. It now not only applies to the genuinely chiral modes found in gyrotropic solids such as quartz, distinguished by their linear wavevector shifts [5,6,7], but also extends to phonons for which the vibrational pattern and the wavevector are co-planar [2,3] as well as circularly-polarized states of zero wavevector [8,9].

Chiral vibrational modes offer a diverse range of possible applications [3], and one area that has garnered significant interest is their ability to manipulate the magnetic properties of a material [10,11]. Central to this domain are coherent states of circularly-polarized polar modes due to their inherent magnetic moment [12,13]. Unlike true chiral phonons [5,6,7] and valley-modes [2,3], the modes of interest here are degenerate. The ions' circular motion arises from the interaction with circularly polarized electromagnetic radiation, which typically generates magnetization due to the inverse Faraday effect [14,15,16]. The experimental data suggests that the intrinsic magnetic moment of the vibrations is too weak to account for the light-induced magnetization and, moreover, that this Faraday magnetization is too small to explain the magnetic effects of the phonons [17]. Here, we show that the root of magnetophononics is not the vibrations themselves, but their coupling to the electrons' spin and orbital degrees of freedom, leading to an ion-induced non-Maxwellian electron field that breaks time-reversal invariance and is effectively larger than the Maxwellian magnetization by several orders of magnitude. We further show that the phonon chirality plays no role in defining the non-Maxwellian field, the behavior of which relies on the angular momentum but not on the wavevector of the vibrations.

Consider a non-magnetic material that belongs to the cubic system and is described by the Hamiltonian

$$\hat{H} = \frac{\hat{\mathbf{P}}^2}{2\bar{m}} + \frac{\bar{m}}{2}\Omega_0^2 \hat{\mathbf{Q}}^2 + \hat{H}_e + \hat{V} \qquad (1)$$

involving a doubly-degenerate, transverse-optical (TO) mode of frequency $\Omega_0$ (the discussion is also applicable to uniaxial substances). Here, $\hat{\mathbf{Q}}$ and $\hat{\mathbf{P}}$ are the phonon coordinate and associated momentum operators, which transform as vectors, $\bar{m}$ is the phonon effective mass [18], $\hat{H}_e$ is the Hamiltonian of the electron subsystem and $\hat{V} = \hat{\mathbf{Q}} \cdot \hat{\mathbf{\Xi}}$ represents the electron-phonon interaction [19]; $\hat{\mathbf{\Xi}} = (\hat{\Xi}_x, \hat{\Xi}_y, \hat{\Xi}_z)$ is a vector operator that depends only on electron variables. Assuming that the ion motion is driven by a classical electric field of frequency $\Omega$ that is polarized in the *xy* plane, and treating the phonon field as classical, we write

$$\hat{V}(t) = \left(Q_R \hat{\Xi}^- + Q_L \hat{\Xi}^+\right) e^{i\Omega t} + \left(Q_R^* \hat{\Xi}^+ + Q_L^* \hat{\Xi}^-\right) e^{-i\Omega t} \quad . \qquad (2)$$

Here, $Q_R = (Q_x + iQ_y)/\sqrt{2}$ and $Q_L = (Q_x - iQ_y)/\sqrt{2}$ are, respectively, the right and left circularly polarized amplitudes and $\hat{\Xi}^\pm = (\hat{\Xi}_x \pm i\hat{\Xi}_y)/\sqrt{2}$. Within the conventional framework of quantum perturbation theory, $\hat{V}(t)$ introduces time-dependent corrections involving both harmonics of $\Omega$ as well as static terms to any given eigenfunction of $\hat{H}_e$. The second-order correction has time-dependent terms at twice the driving frequency as well as the lowest-order static correction

$$\Psi_p^{(2)} = \frac{1}{\hbar^2} \sum_{n,m} \left[ \frac{\left(Q_R^* \hat{\Xi}_{nm}^+ + Q_L^* \hat{\Xi}_{nm}^-\right)\left(Q_R \hat{\Xi}_{mp}^- + Q_L \hat{\Xi}_{mp}^+\right)}{\omega_{np}(\omega_{mp} - \Omega)} + \frac{\left(Q_R \hat{\Xi}_{nm}^- + Q_L \hat{\Xi}_{nm}^+\right)\left(Q_R^* \hat{\Xi}_{mp}^+ + Q_L^* \hat{\Xi}_{mp}^-\right)}{\omega_{np}(\omega_{mp} + \Omega)} \right] \Psi_n^{(0)}, \qquad (3)$$

where $\hat{H}_e \Psi_n^{(0)} = \hbar \omega_n \Psi_n^{(0)}$ and $\omega_{mp} = \omega_m - \omega_p$. This expression contains a symmetric term and, central to magnetophononics, the antisymmetric contribution

$$\Psi_{p,\mathrm{AS}}^{(2)} = \frac{\Omega}{\hbar^2} \left(|Q_R|^2 - |Q_L|^2\right) \sum_{n,m} \frac{(\hat{\Xi}_{nm}^+ \hat{\Xi}_{mp}^- - \hat{\Xi}_{nm}^- \hat{\Xi}_{mp}^+)}{\omega_{np}(\omega_{mp}^2 - \Omega^2)} \Psi_n^{(0)}, \qquad (4)$$

which vanishes at $\Omega = 0$. Had we considered instead of the electron-phonon the electron-radiation interaction $\mathbf{E} \cdot \hat{\mathbf{d}}$, where $\mathbf{E}(t)$ is a classical electric field and $\hat{\mathbf{d}}$ stands for the electron dipole-moment operator, the perturbation expansion would be the standard one of nonlinear optics, with the second-order symmetric correction to the ground state leading to second harmonic generation and optical rectification in non-centrosymmetric materials [20], and the antisymmetric contribution to the inverse Faraday effect [14]. Building on this analogy, it follows that the circular motion of the ions will induce a magnetization along the z direction, given by

$$M_z = \frac{\Omega}{\hbar^2 v_c} \left(|Q_R|^2 - |Q_L|^2\right) S(\Omega) \qquad (5)$$

with

$$S(\Omega) = \sum_{n,m} \frac{(\hat{\Xi}_{gm}^+ \hat{\Xi}_{mn}^- - \hat{\Xi}_{gm}^- \hat{\Xi}_{mn}^+) \langle n | m_z | 0 \rangle}{\omega_{ng}(\omega_{mg}^2 - \Omega^2)} + c.c. \qquad (6)$$

Here, $v_c$ is the unit cell volume, $\hat{m}_z$ is the z component of the electron magnetic dipole operator, c.c. denotes the complex conjugate, and we have assumed that the ground state $|0\rangle$ is not degenerate and the temperature is $T = 0$. Since $Q$ and $E$ are related through [12]

$$Q_{x,y} = \sqrt{\frac{4\pi v_c / \overline{m}}{(\varepsilon_0 - \varepsilon_\infty)}} \frac{\chi^{(1)}(\Omega)}{\Omega_0} E_{x,y} \quad , \tag{7}$$

where $\varepsilon_0$ ($\varepsilon_\infty$) is the low (high) frequency permittivity and $\chi^{(1)}(\Omega)$ is the linear optical susceptibility, the well-known relationship between the inverse Faraday and the conventional Faraday effect [15] can be used to obtain the angle of rotation $\theta_F$ of a linearly-polarized beam in the presence of a magnetic field **B** oriented along $z$

$$\theta_F = \frac{16\pi^2 (\Omega/\Omega_0)^2}{\overline{m}\hbar^2 n_r(\Omega) c} \frac{\left|\chi^{(1)}(\Omega)\right|^2 \ell |\mathbf{B}|}{(\varepsilon_0 - \varepsilon_\infty)} S(\Omega) \quad . \tag{8}$$

Here, $\ell$ is the beam path length, $c$ is the speed of light and $n_r$ is the refractive index. The ratio $\theta_F / \ell |\mathbf{B}|$ gives the ($\Omega$-dependent) Verdet constant. When setting $T = 0$, the paramagnetic term, which is usually dominant in substances featuring ions with unfilled orbitals, is effectively eliminated from consideration. Provided the Zeeman splittings are small compared to $k_B T$ ($k_B$ is Boltzmann constant), this term manifests itself as a $1/T$ contribution to the Verdet constant [15].

It is instructive to compare Eqs. (5) and (8) with the corresponding expressions for the direct coupling of the magnetic field to the phonon magnetic moment, which can be written as $g_P \overline{m}(\mathbf{Q} \times \dot{\mathbf{Q}}) \cdot \mathbf{B}$, where $g_P$ is the phonon gyromagnetic ratio [12,13]. This interaction gives

$$\theta'_F = \frac{16\pi^2 g_P (\Omega/\Omega_0)^2}{n_r(\Omega) c} \frac{\left|\chi^{(1)}(\Omega)\right|^2 \ell |\mathbf{B}|}{(\varepsilon_0 - \varepsilon_\infty)} \tag{9}$$

And, thus, $M'_z = (g_P \Omega \overline{m} / v_c)\left(|Q_R|^2 - |Q_L|^2\right)$. For insulators, one gets the crude estimate $M_z / M'_z = \theta_F / \theta'_F \sim v_c^{2/3} \Xi_{ave}^2 / E_G^2 \sim 10^2 - 10^3$ using typical values of the TO deformation potential [21] and a gap of $E_G = 1$ eV. This ratio is also the extent by which $\partial \Omega_0 / \partial B$ is greater for the indirect, electron-mediated mechanism of phonon-magnetic field coupling. In its many forms, this mechanism accounts for the large magnetic-field dependence of phonon frequencies observed in rare-earth compounds like CeF3, which closely follow the average magnetic moment of the rare-earth ions [22,23], topological semimetals [24], incipient ferroelectrics [25] and polar antiferromagnets [26].

Equation (4) can be seen as arising from an effective, static perturbation $\hat{V}_{\text{eff}}$ that acts on the electrons, with matrix elements

$$\langle \Psi_n^{(0)} | \hat{V}_{\text{eff}} | \Psi_p^{(0)} \rangle = \left( \mathbf{Q}(\Omega) \times \mathbf{Q}^*(\Omega) \right) \cdot \sum_m \frac{\Omega \left( \vec{\Xi}_{nm} \times \vec{\Xi}_{mp} \right)}{\hbar(\omega_{mp}^2 - \Omega^2)} \quad . \tag{10}$$

The behavior of vector products closely mimics that of a magnetic field, for they are even under inversion and odd under time reversal. Consequently, the product $\mathbf{Q} \times \mathbf{Q}^*$ displays characteristics of a non-Maxwellian magnetic-esque field, inducing the splitting of Kramers doublets and modifying the electron energy spectrum in a manner reminiscent of an actual field. Unlike true magnetic fields, however, the non-Maxwellian field is undetectable outside the material.

Understanding the time reversal properties of $\hat{V}_{\text{eff}}$ is easiest in the absence of spin-orbit coupling. In such instances, it suffices to consider the orbital component of the electron magnetic dipole $\hat{\mathbf{m}}_{\text{L}} = -e(\hat{\mathbf{r}} \times \hat{\mathbf{p}})/2m_e c$. Since $\hat{\Xi}$ transform like $\hat{\mathbf{r}}$, the exact relationship

$$\langle s | \hat{\mathbf{m}}_{\text{L}} | t \rangle = i \frac{e}{2c} \sum_m \omega_{mt} (\hat{\mathbf{r}}_{sm} \times \hat{\mathbf{r}}_{mt}) \tag{11}$$

can be used to obtain the approximate expression

$$\hat{V}_{\text{eff}} \approx -(|Q_R|^2 - |Q_L|^2) \frac{2\hbar^2 \Omega c}{e(\Delta^2 - \hbar^2 \Omega^2)\Delta} \frac{\Xi_{\text{ave}}^2}{r_{\text{ave}}^2} \hat{\mathbf{m}}_{\text{L}} \cdot \mathbf{e}_z \quad , \tag{12}$$

which applies when transitions between two particular electron states or two bands dominate. Here $\Delta$ is the energy separation between the two states while $\Xi_{ave}^2$ and $r_{ave}^2$ are average values of the corresponding matrix elements. Within this approximation, the factor multiplying $\hat{\mathbf{m}}_L \cdot \mathbf{e}_z$ plays the role of an effective magnetic field oriented along *z*. An order-of-magnitude estimate indicates that the ratio between this field and the magnetization, Eq. (5), is $\approx \alpha^{-2}$, where $\alpha \approx 1/137$ is the fine structure constant. Drawing from the analogy with $\mathbf{E} \cdot \hat{\mathbf{d}}$ coupling [14], one can further show that the effective interaction for paramagnetic ions is of the form

$$(|Q_R|^2 - |Q_L|^2) A_Q(\Omega) S_z \quad , \tag{13}$$

where $A_Q$ depends linearly on the spin-orbit interaction (this presupposes that the orbital angular momentum of the ground state multiplet is quenched). As for the spinless case, the magnitude of $A_Q$ is such that the splittings due to $\hat{V}_{eff}$ are on the order of $\alpha^{-2}$ times larger than $4\pi\mu_B M_z$; $\mu_B$ is the Bohr magneton.

The result that $\hat{V}_{eff}$-induced splittings surpass those of the light-induced magnetization by a factor of approximately $\alpha^{-2}$ is a key outcome of this work. This becomes readily understandable when considering that the magnetic field strength needed to produce **M** is $\mathbf{M}/\chi_M$ where $\chi_M$ is the static magnetic susceptibility, the values of which typically fall around $\alpha^2$, give or take one order of magnitude in variation. The estimated values of the effective field are quite large. Taking, as previously, $E_G = 1$ eV and using representative values for the electron-phonon coupling [21], one gets $|\mathbf{M}|\alpha^{-2} = 50\text{-}100$ T for $|\mathbf{Q}| = 0.2$ Å and a driving frequency of 10 THz. Needless to say, the actual $\hat{V}_{eff}$ splittings and those due to $\mathbf{M}/\chi_M$ may differ considerably and, moreover, our continuous wave approach requires refinement to better suit the analysis of pulsed experiments.

Even though $\hat{V}_{eff}$ is associated with a non-Maxwellian field and, as such, hidden to the outside world, its impact inside the material has been well documented across experiments encompassing the polarization rotation of a linearly-polarized probe beam (optical Faraday effect [17]) and the phonon-induced generation of coherent magnons [10]. Concerning the former, it should now be clear that the field responsible for the probe polarization rotation is not the Maxwellian field derived from the magnetization, namely, $\mathbf{B} = 4\pi\mathbf{M}$, as sometimes assumed, but the much larger non-Maxwellian field associated with $\hat{V}_{eff}$. Additionally, Eq. (13) indicates that the sudden onset of coherent phonon oscillations can coherently drive a magnetic precession provided the spin magnetization axis is perpendicular to the light-induced magnetic-like field. In this regard, it is worth noting that the experiments on $ErFeO_3$ are consistent with our findings; both the coupling related to the magnetization, as per Eq. (5), and the magnetic moment attributed solely to ion motion fail to account for the notably larger values of the effective magnetic field responsible for driving the magnon oscillations [10].

The preceding discussion centered on the behavior of the antisymmetric component of the nonlinear susceptibility $\chi^{(2)}$ at far infrared frequencies, close to those of polar modes. Nevertheless, these findings extend their significance to Faraday-related experiments performed at higher frequencies [27,28], where the phonons play no role. As alluded to earlier, the formulas governing light-induced magnetization, Eq. (5), and the effective static perturbation, Eq. (10), align precisely with standard nonlinear optics equations when $\mathbf{Q}$ is exchanged for the electric field $\mathbf{E}$ and $\hat{\Xi}$ for the electron dipole operator $\mathbf{d}$ [15]. Whether at infrared or higher frequencies, the fact that Faraday-related effects depend on the quadratic combination $\mathbf{Q} \times \mathbf{Q}^*$ (or $\mathbf{E} \times \mathbf{E}^*$) indicate that the chiral properties inherent in circularly-polarized electromagnetic radiation, which are transferred to the material, become entirely inconsequential. Furthermore, since the group velocity of optical modes vanishes at the center of the Brillouin zone and the chiral pitch is orders-of-magnitude larger than the lattice parameter, it becomes evident that any effects stemming from the phonon wavevector will be negligible. This strongly suggests that chirality is not a factor in the generation of magnetic-like fields, as commonly proposed [29,30,31,32].

To summarize, our findings highlight the pivotal role of electron-phonon coupling in elucidating the magnetic effects arising from the excitation of polar phonons. The circular motion of ions induces a nonreciprocal and achiral static perturbation. Its impact on the electrons mirrors that of a magnetic field, reaching magnitudes we estimate can be as large as 100 T, markedly surpassing the fields generated by the ion motion alone. These results promise magnetic fields that exceed current experimental limitations, achievable within a table-top setup, which could pave the way for delving into novel optical phenomena, advancing spin control crucial for quantum computing, and investigating quantum-Hall-related effects dependent on broken time reversal symmetry.

.